# A multi-dimensional analysis of usage counts, Mendeley readership, and citations for journal and conference papers

Wencan Tian[1], Zhichao Fang[2,3], Xianwen Wang[1], Rodrigo Costas[3,4]
1.  WISE Lab, Institute of Science of Science and S&T Management, Dalian University of Technology, Dalian, China
2.  School of Information Resource Management, Renmin University of China, Beijing, China
3.  Centre for Science and Technology Studies (CWTS), Leiden University, Leiden, The Netherlands
4.  DSI-NRF Centre of Excellence in Scientometrics and Science, Technology and Innovation Policy, Stellenbosch University, Stellenbosch, South Africa

* Corresponding author: xianwenwang@dlut.edu.cn

**Abstract:** This study analyzed 16,799 journal papers and 98,773 conference papers published by IEEE Xplore in 2016 to investigate the relationships among usage counts, Mendeley readership, and citations through descriptive, regression, and mediation analyses. Differences in the relationship among these metrics between journal and conference papers are also studied. Results showed that there is no significant difference between journal and conference papers in the distribution patterns and accumulation rates of the three metrics. However, the correlation coefficients of the interrelationships between the three metrics were lower in conference papers compared to journal papers. Secondly, funding, international collaboration, and open access are positively associated with all three metrics, except for the case of funding on the usage metrics of conference papers. Furthermore, early Mendeley readership is a better predictor of citations than early usage counts and performs better for journal papers. Finally, we reveal that early Mendeley readership partially mediates between early usage counts and citation counts in the journal and conference papers. The main difference is that conference papers rely more on the direct effect of early usage counts on citations. This study contributes to expanding the existing knowledge on the relationships among usage counts, Mendeley readership, and citations in journal and conference papers, providing new insights into the relationship between the three metrics through mediation analysis.

# 1 Introduction

The evaluation of scholarly articles has always been a fundamental element in research evaluation, being essential in evaluating researchers for career advancement, project applications, or just their academic reputation (Breitzman, 2021). Traditionally, citation-based metrics have been used to evaluate the impact of individual publications (Waltman, 2016). However, these metrics have increasingly come under scrutiny due to their complex motives and citation delay phenomena (Cui et al., 2023; Khan & Younas, 2017; Wang et al., 2016), as well as other more fundamental issues like their meaning or conceptual ambiguity (Hicks et al., 2015).

In this context, altmetrics garnered considerable attention in recent years (Costas et al., 2015; Chi et al., 2019; Chen et al., 2020; Geng et al., 2022) since they offered potentially "alternative" evaluation metrics for scientific publications by capturing social media attention, online discussions, and other forms of non-traditional impact, thus potentially supplementing traditional citation-based metrics (Erdt et al., 2016; Sugimoto et al., 2017). Altmetrics, or social media metrics (Haustein et al., 2016), were therefore expected to provide faster metrics for research evaluation and, to some extent, mitigate the citation delay issue (Wang et al., 2016; Khan & Younas, 2017). However, previous studies have demonstrated the unsuitability of social media metrics, mostly based on social media events (e.g., Twitter, Facebook), for research evaluation or even predicting citations (Haustein et al., 2014; Fang et al., 2020; Zahedi & Costas, 2018). The most notable exception is the case of Mendeley readership, the only altmetric source to show a moderate correlation with citations (Fang et al., 2020; Zahedi et al., 2014) and similar distribution properties as citations (Costas et al., 2017). Another important source of metrics for scientific publications is *usage metrics*, which have been researched for even longer than altmetrics (Glanzel & Gorraiz, 2015) and also provide relevant data on how individual scientific publications are being viewed or downloaded (typically referring to the number of HTML views or PDF downloads) by different users.

Mendeley readership and usage counts, together with the number of (re)tweets of articles, are the metrics that have been studied more often in their relationships with citations (Wang et al., 2014; Mohammadi et al., 2015; Fang et al., 2022, 2021). However, a combined study of usage counts, Mendeley readership and citations, particularly paying attention to the potential mediation effects among them, have never been done before. This study aims at better understanding the relationship between the first two (usage counts and Mendeley readership) and citations, and comprehensively analyzes journal and conference papers' similarities and differences.

## 1.1 Relationship between usage counts and citations

In the extant literature on the interplay between usage counts and citations, most research focused on exploring the correlation between these two metrics. Such inquiries aimed at developing novel academic evaluation metrics (such as the "usage impact factor" (Bollen & van de Sompel, 2008; Schloegl & Gorraiz, 2010), the "usage immediacy index" (Rowlands & Nicholas, 2007), or the "download immediacy index" (Wan et al., 2010)) to assess individual researchers, academic journals, and institutions' research capabilities. Scholars have conducted a considerable amount of related research on different disciplines, which yielded different conclusions. Most previous studies pointed to a positive correlation between usage and citation counts (Bollen & van de Sompel, 2008; Bollen et al., 2003; Chi & Glänzel, 2018; McGillivray & Astell, 2019). For instance, in the field of chemistry, the usage and citation data on the Web of

Science platform showed a moderate correlation (Chi et al., 2019). Lippi & Favaloro (2013) identified a strong correlation between article rankings based on downloads and the relative number of ScienceDirect citations. Chi & Glanzel (2017) conducted a comparative analysis of usage and citation data of articles published in Web of Science from three countries, Belgium, Israel, and Iran, and found that citations and usage counts are significantly correlated, particularly in the social sciences. Furthermore, Ding et al. (2021) verified the Granger causality relationship between usage and citation counts in more than 7,000 articles published in *The Lancet*.

In addition to examining the correlation between usage and citation counts at the article level, some scholars have also investigated the relationship between these two metrics at the journal level, finding that the correlation is stronger at the journal level (Vaughan et al., 2017; Guerrero-Bote & Moya-Anegón, 2014; Schloegl & Gorraiz, 2010). Furthermore, papers published in non-English-language journals exhibit a higher correlation between usage counts and citations than those published in English-language journals (Guerrero-Bote & Moya-Anegón, 2014). In addition, to address the research gap in exploring the relationship between usage counts and citations using Chinese databases, Vaughan et al. (2017) used a sample of 150 journals from the China Academic Journal Network Publishing Database and demonstrated a strong correlation between usage counts and citations at the journal level.

**1.2 Relationship between Mendeley readership and citations**
The relationship between Mendeley readership and citations has been investigated in several studies. It is widely acknowledged within the scientific community that a positive correlation exists between these two metrics (Zahedi et al., 2017; Zahedi & Haustein, 2018). For example, Li et al. (2012) analyzed 1,613 articles published in *Nature* and *Science* and reported a significant correlation between Mendeley readership and WoS citations. In another study, Thelwall (2017) compared the correlation between Mendeley readership and Scopus citations for journal articles in 325 narrow Scopus fields and found a strong positive correlation in most fields, with an average correlation coefficient of 0.671, and a higher correlation in social sciences than in humanities (Mohammadi & Thelwall, 2014). Notably, the proportion of researchers who actually read the articles saved in Mendeley may affect the reliability of the aforementioned results. To address this issue, Mohammadi et al. (2016) surveyed 860 Mendeley users and found that 55% of users had created a personal library in Mendeley and claimed to have read or planned to read at least half of the academic articles included in their library. Moreover, 85% of respondents indicated that using Mendeley had facilitated their future citation work.

**1.3 Relationship between usage counts and Mendeley readership**
Compared to the two types of relationship research mentioned above, research on the relationship between usage counts and Mendeley readership has been relatively limited. Existing research suggests that the correlation between these two metrics is generally low to moderate. For instance, Schloegl et al. (2014) studied articles published in the *Journal of Strategic Information Systems* and *Information and Management* and discovered correlation coefficients between the two metrics of 0.73 and 0.66, respectively. In a separate investigation, Thelwall & Kousha (2017) revealed that the correlation coefficient between the two metrics was distributed between 0.2 and 0.4. Additionally, Wang et al. (2020) found that the correlation coefficient between usage counts and Mendeley readership did not exhibit a considerable difference between preprints and non-open access papers, with both coefficients distributed between 0.18 and 0.52.

**1.4 Objectives**

Based on the literature review presented above, it is evident that most studies examining the correlation among the three metrics have primarily utilized data derived from journal articles, while limited attention has been given to conference papers. Additionally, there is a paucity of deeper path analysis or interaction mechanism analysis of the relationship between the three metrics. Given these gaps in the literature, this paper employs multiple perspectives, including descriptive, regression, and mediation analyses, to investigate the similarities and differences in the relationship among the three metrics in both journal and conference papers. The specific research questions in this study are as follows.

RQ1. Are the distribution patterns, correlations, and accumulation rates of usage counts, Mendeley readership, and citations in journal papers and conference papers similar? Furthermore, what is the relationship between funding, international collaboration, and open access with these three metrics? Does this relationship differ between journal and conference papers?

RQ2. Which metric, early usage counts or early Mendeley readership, performs better in predicting the future academic citation impact of articles?

RQ3. Does early Mendeley readership mediate the relationship between early usage counts and later citation impact? And if so, how does this mediating effect differ between journal papers and conference papers?

## 2 Data and Methods

**2.1 Dataset**

Data were obtained from articles published in the IEEE Xplore database in 2016, including 16,799 journal papers (published in 134 distinct journals) and 98,773 conference papers. The IEEE Xplore database is a professional electronic, electrical, and computer engineering database (Khan & Younas, 2017; Tian et al., 2019). It encompasses a rich collection of journal articles and conference proceedings related to these fields and has been providing monthly usage data (total number of HTML views and PDF downloads) for each item since 2011 (Breitzman, 2021). Using web crawling techniques, we obtained the annual usage data for the articles mentioned above between 2016 and 2020, including 839,995 usage data points generated by journal papers and 493,865 usage data points generated by conference papers. In addition, we retrieved the annual Mendeley readership and citation counts for the publications above between

2016 and 2020 from the CWTS in-house database. The Mendeley readership data were collected by using the Mendeley API on a yearly basis (in July) since 2016, while the annual citation counts of the publications were obtained from the Dimensions database. As a result, for each publication in our dataset, we calculated the usage counts, Mendeley readership, and citations for each year within the observation time window of 2016 - 2020.

## 2.2 Introduction of variables

To address the study's second research question, which investigates the ability of early usage counts and early Mendeley readership to predict future citation counts of an article, *early usage counts* and *early Mendeley readership* are operationalized as the cumulative counts within the first two years after publication, whereas *future citations* are measured by the cumulative counts within five years after publication. Additionally, the study controlled for several potential confounding variables (Cui et al., 2023), such as whether the article received funding, involved international collaboration, was open access, and the journal impact factor (JIF). Funding, international collaboration, and open-access information were obtained directly from the Dimensions database hosted at CWTS. The JIF information was obtained from the Journal Citation Reports (JCR) 2016 provided by the Web of Science. Table 1 provides a detailed description of the variables.

**Table 1** Variable description

| Variable | Description |
|---|---|
| Citations | It is the dependent variable in all models and represents the total number of citations an article has received within the five years since publication. |
| Early usage counts | Usage counts of an article in the publication year and the following year. |
| Early Mendeley readership | Number of Mendeley readers of an article in the publication year and the following year. |
| Funding | Set to 1 if an article is supported for funding; otherwise 0. |
| International collaboration | Set to 1 if an article developed international collaboration; otherwise, mark 0. |
| Open access | Set to 1 if an article is open access (as in Dimensions); otherwise 0. |
| JIF | Journal impact factor of the journal in which an article was published. It was retrieved from the Journal Citation Reports for the year 2016. |

## 2.3 Mediation effects

Mediation analysis is a powerful research tool in social sciences that facilitates a better comprehension of the fundamental mechanisms through which variables interact with each other. The mediation effect analysis was initially developed in psychology (Baron & Kenny, 1986) and gradually extended to the fields of management and economics (Raguseo et al., 2021; Singh et al., 2023). More recently, this method has also been applied in scientometrics. For instance, Álvarez-Bornstein & Bordons (2021) examined the mediating effects of journal quartile, collaboration type, and the number of references on the relationship between funding and article impact. They found that the presence and magnitude of this effect varied by discipline. Ebrahimy et al. (2016) assessed whether three article-level metrics provided by PLOS - save, discussion, and recommendation metrics - were mediators between the visibility and citations of biomedical articles. Their findings indicated that only the save metric had a positive mediating effect in the relationship between visibility and citations, while recommendation metrics had no impact on this relationship. They also found that discussion metrics played a negative mediating role in this relationship between visibility and citations. Using articles from the biomedical field as the data sample, Vilchez-Roman and Vara-Horna (2021) used usage frequency as a mediating variable to explore the effect of social media platforms such as Twitter and Facebook on the citations. The study revealed that while the direct impact of Twitter on citations was negative, the indirect effect through usage frequency was positive and significant.

Building on prior research, this article extends the data sample to the IEEE Xplore 2016 dataset and performs an examination of the mediating mechanisms between usage counts, Mendeley readership, and citations by controlling for relevant variables. Furthermore, a comparative analysis between journal and conference papers is conducted to expand the study's insights.

Mediation analysis can be conducted using two main methods: traditional regression analysis and structural equation modeling (SEM). As illustrated in Fig. 1(a-c), regression analysis involves three steps. First, the coefficient a is estimated by regressing Y on X to assess the total effect. The coefficient b is then estimated by regressing M on X to examine the relationship between the explanatory variable and the mediator variable. Finally, the coefficient c is estimated by regressing Y on M while controlling for X to determine the relationship between the dependent variable and the mediator variable, and the estimated coefficient a' can also be obtained. When all coefficients a, b, and c are significant, a' can be used to determine the presence and extent of mediation effects. Specifically, if a' is significant and equals 0, M has a complete mediation effect, while if a' is not significant and does not equal 0, M has a partial mediation effect. Of course, the existence of partial mediation indicates a decrease in the direct effect of path a. To determine the magnitude of this decrease, further validation is required using methods such as the Sobel test. However, as shown in Fig. 1(d), SEM allows for the simultaneous estimation of all model parameters, which can address the large standard errors and inaccurate parameter estimates that can occur when using regression analysis (Iacobucci et al., 2007), making it a superior method for conducting mediation analysis. Therefore, this study employs SEM for mediation analysis and follows the methods proposed by Baron and Kenny (1986) and Zhao et al. (2010) to test mediation effects. The *medsem* program developed by Mehmetoglu in 2018 is used to calculate direct and indirect effects (Mehmetoglu, 2018).

Stata 17.0 was used as the analysis tool.

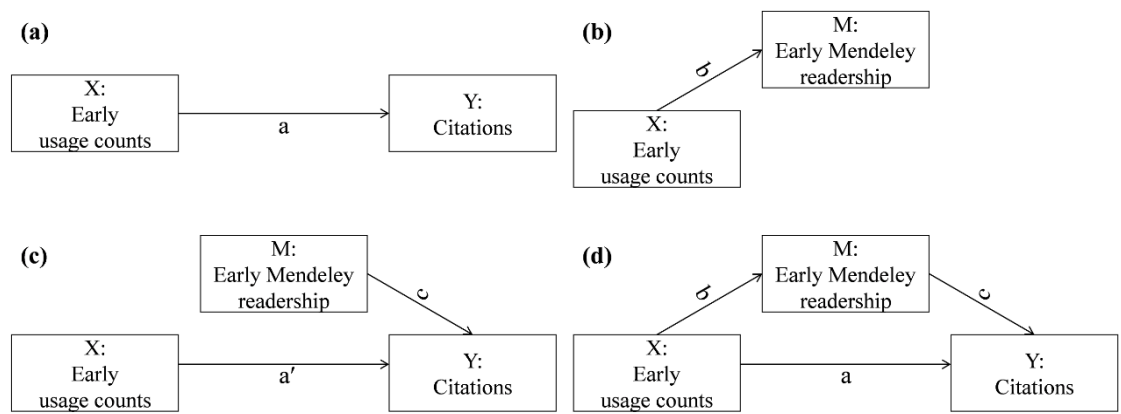

**Fig. 1** Diagrammatic illustration of two mediation analysis methods: regression analysis (a-c) and structural equation modeling (d)

The mediation analysis using SEM follows two steps. Firstly, the model is fitted to estimate the coefficients of the direct effect and the mediation effect of early usage counts on citations. If both $X \to M$ and $M \to Y$ are significant, there is a mediation effect, and the research should proceed to step two. If neither coefficient is significant, there is no mediation effect, and the research should be terminated. Secondly, the size of the mediation effect relative to the direct effect is evaluated using the Sobel test, the Delta test, and the Monte Carlo simulation test. If the Z-value based on these tests is significant and the coefficient $X \to Y$ is not significant, early usage counts have a full mediation effect on citations, indicating that early usage counts cannot directly affect citations but only have an indirect effect through early Mendeley readership. If both the Z-value and coefficient $X \to Y$ are significant, early usage counts have both a direct effect on citations and an indirect effect through early Mendeley readership. Finally, the final estimation results are organized and reported, with three possible outcomes: no mediation, partial mediation, or full mediation.

## 3 Results

### 3.1 Descriptive analysis

As shown in Fig. 2, we present the distribution of usage counts, Mendeley readership, and citations (statistical information about the three metrics can be found in Table 6 in the Appendix). It is worth noting that the usage counts and Mendeley readership mentioned in Section 3.1 are cumulative totals in the five years since the article's publication. It can be observed that the usage counts of both journal articles and conference papers exhibit a similar trend, with an initial increase followed by a subsequent decline. In contrast, citations follow a power-law distribution. Specifically, the distribution of usage counts for journal articles is mainly concentrated between 200 and 400, with only a small proportion of articles exceeding 1,000. For conference papers, the distribution of usage counts is focused between 50 and 150, with a very limited number of articles surpassing 500. Furthermore, the distribution of Mendeley readership exhibits a significant difference between journal and conference papers, with the former following a pattern similar to that of usage counts and the latter following a pattern comparable to that of citations.

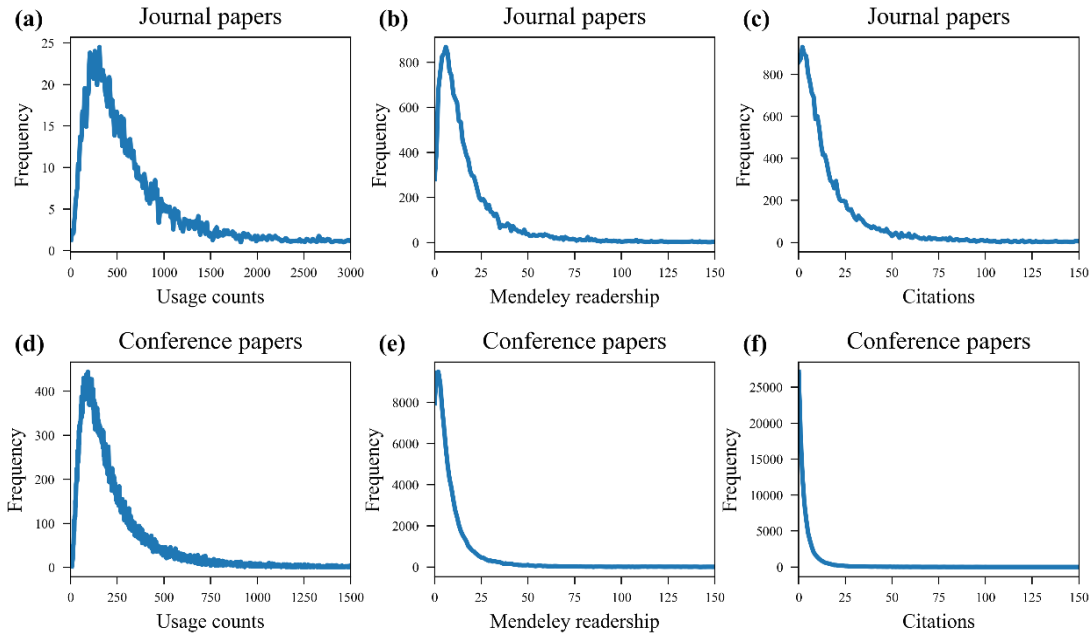

**Fig. 2** Distribution of usage counts, Mendeley readership, and citations in journal (subfigures a-c) and conference (subfigures b-d) papers

Fig. 3 illustrates the pairwise correlations between the metrics of usage counts, Mendeley readership, and citations. Spearman correlation coefficients are reported in the subfigures. It can be observed that these three metrics have significant positive correlations, with the overall correlation coefficients in conference papers being lower than those in journal papers. Specifically, the correlations between the metrics in journal papers are all higher than 0.68, with the strongest correlation observed between usage counts and citations ($r = 0.722$). In contrast, the correlation coefficients for conference papers are relatively lower, with the highest correlation coefficient observed between usage counts and Mendeley readership ($r = 0.586$). This suggests that both usage counts and Mendeley readership are relevant predictors of citations, regardless of publication type. Additionally, the correlations among these three metrics vary depending on the publication type.

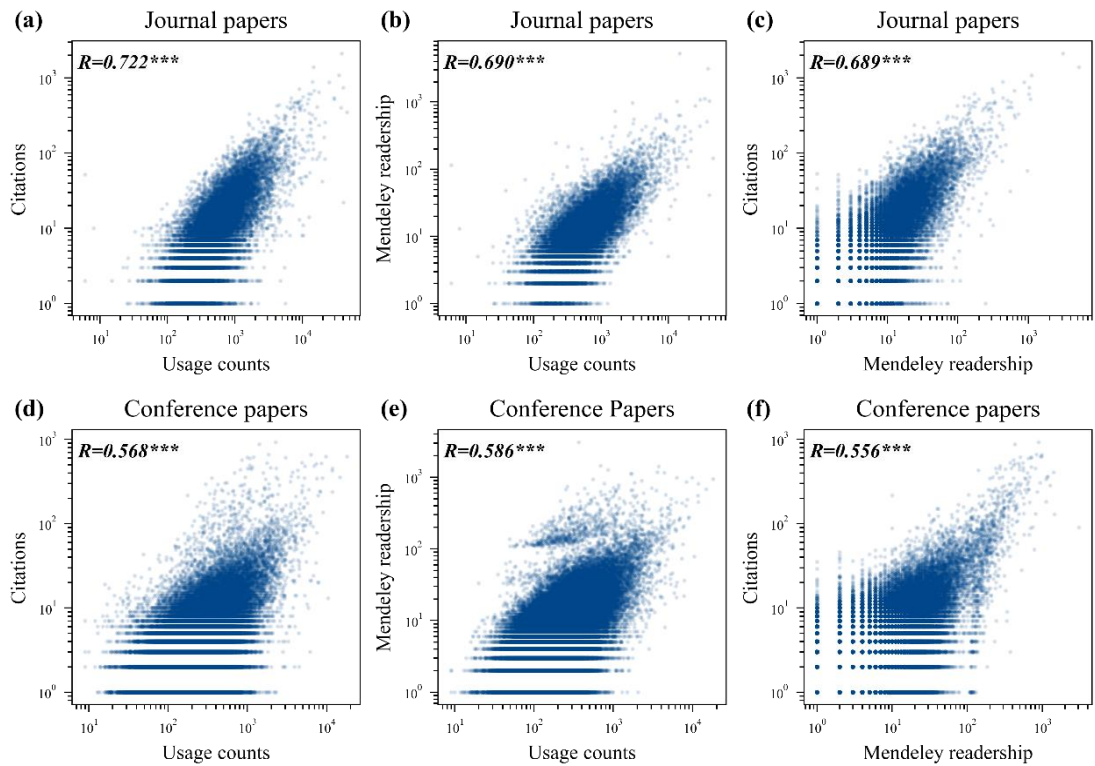

**Fig. 3** Pairwise correlation between usage counts, Mendeley readership, and citations for journal (subfigures a-c) and conference (subfigures b-d) papers. *** indicates that the correlation coefficient is statistically significant at the 0.001 level, same below.

We conducted further investigation into the temporal dynamics of usage counts, Mendeley readership, and citations during the 5-year period after publication. As depicted in Fig. 4 (a-b), the usage counts of journal articles reached their peak in the first year, subsequently declining rapidly, whereas for conference papers, they peaked in the second year before decreasing. On the other hand, Mendeley readership experienced its highest point in the second year, followed by a downward trend that was more conspicuous for conference papers. Meanwhile, citations tended to stabilize from the third year onward. As demonstrated in Fig. 4 (c-d), irrespective of the publication type, the dissemination speed of these metrics seems to follow the order of "usage counts > Mendeley readership > citations". Notably, the dissemination speed of the Mendeley readership is nearly indistinguishable from that of citations.

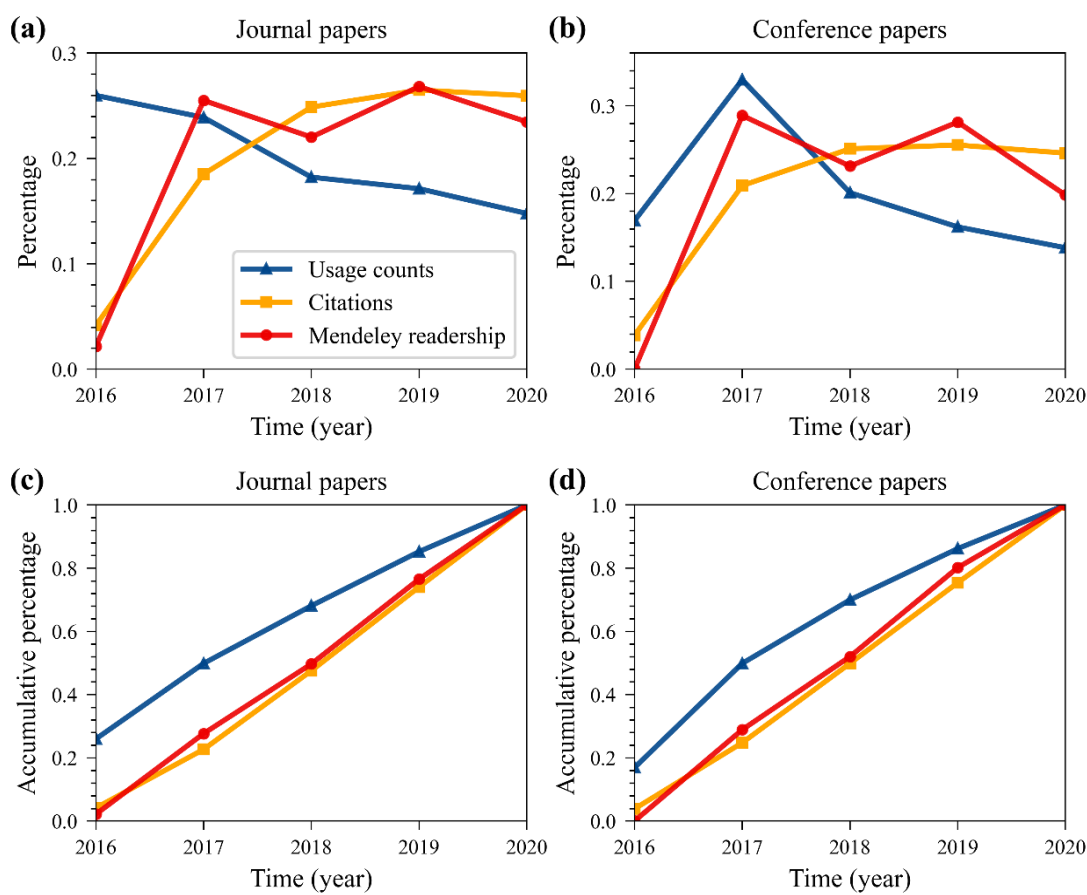

**Fig. 4** Changes in the usage counts, Mendeley readership, and citations of journal papers (subfigures a and c) and conference papers (subfigures b and d) over time.

Finally, we explored the relationship between other bibliometric indicators – the presence of funding, international collaboration, and open access - and usage counts, Mendeley readership, and citations (e.g., do funded articles have higher usage counts, Mendeley readership, or citations compared to unfunded articles?). In this exploration, we initially grouped the journal/conference articles based on whether they received funding (or not), following the same grouping approach for publications with/out international collaboration and with/out open access. Subsequently, we aggregated the means of usage counts/Mendeley readership/citations within each group (as shown in the bars in all subfigures in Fig. 5, where the error bars represent 95% confidence intervals). Finally, employing the *statsmodels* package in Python (https://www.statsmodels.org/stable/index.html), we conducted a one-way analysis of variance (ANOVA) test to examine whether a statistically significant difference exists in the usage counts/Mendeley readership/citations between the two groups (the statistical test result is marked in the upper right corner of all subfigures in Fig. 5).

As shown in Fig. 5, the findings reveal that funding, international collaboration, and open access are positively associated with all three indicators, except for the insignificant relationship of funding with the usage counts of conference papers. Furthermore, the relationship of open access with the three indicators is greater than that of funding and international collaboration, as previously demonstrated in existing research on the advantages of open access in terms of usage counts and citations (Wang et al., 2015; Holmberg et al., 2020). Additionally, while international collaboration and funding have a similar relationship with citations and Mendeley readership, international collaboration outperforms funding in terms of usage counts, particularly in journal articles. Specifically, in journal articles, international collaboration (*vs.* non-international cooperation) could increase the usage counts of an article by 266 (average), but funding (*vs.* non-funding) could only increase it by 120 (average). It is worth mentioning that more detailed information about the statistics of Fig. 5 can be found in Table 7 in the Appendix.

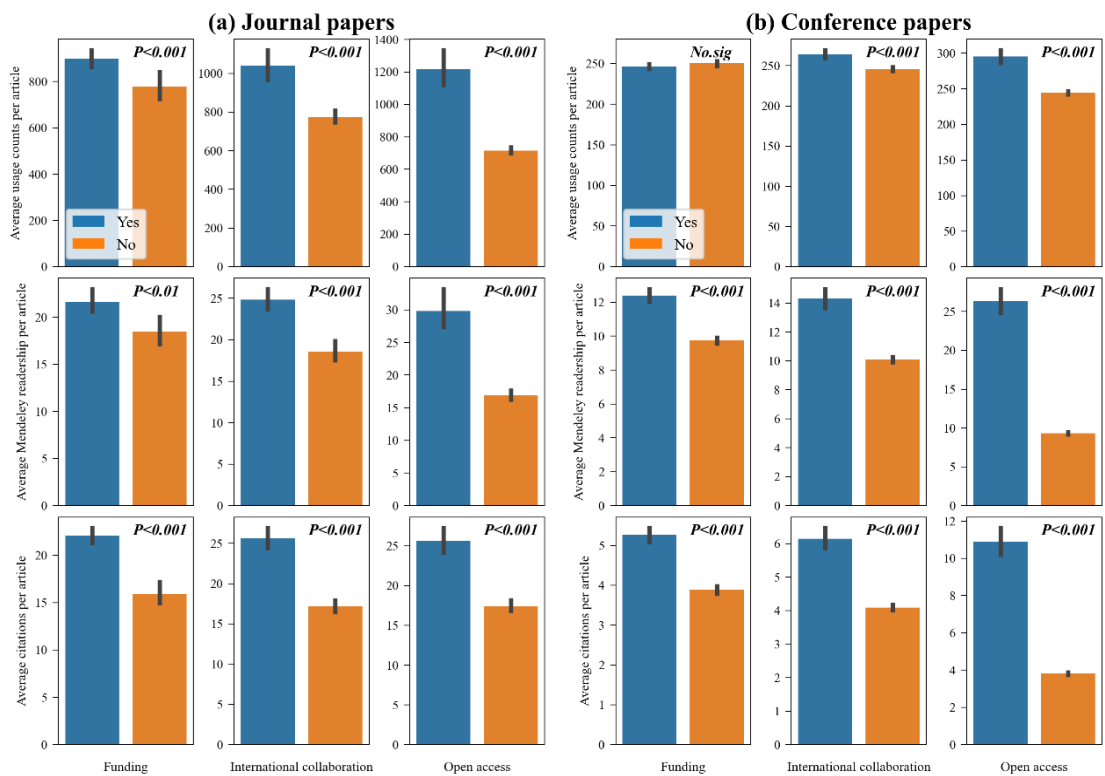

**Fig. 5** Relationship between funding/international collaboration/open access and usage counts/Mendeley readership/citations for (a) journal papers and (b) conference papers. "Yes" means that articles in this group were funded by grants or developed international collaborations or were open access. "No" means the opposite.

For journal papers, we conducted further investigations into the associations between usage counts, Mendeley readership, citations, and JIF at the journal-level (including 134 journals). All three indicators were significantly and positively correlated with JIF (see Fig. 6) at the journal-level. Specifically, the Spearman correlation coefficients between usage counts and JIF, Mendeley readership and JIF, and citations and JIF were 0.570, 0.584, and 0.775, respectively. Notably, the robust positive correlation between citations and JIF suggested that papers published in high-impact journals tended to garner more citations (Cui et al., 2023; Vaughan et al., 2017). Moreover, the moderate correlation between usage counts/Mendeley readership and JIF implied that articles published in high-impact journals were also more likely to attract more readers on average, although with a weaker relationship than in the case of citations. This suggests that the impact of a journal cannot just be captured by their JIF, but that metrics like usage counts or Mendeley readership can contribute to providing a more comprehensive and nuanced reflection of an actual journal's impact.

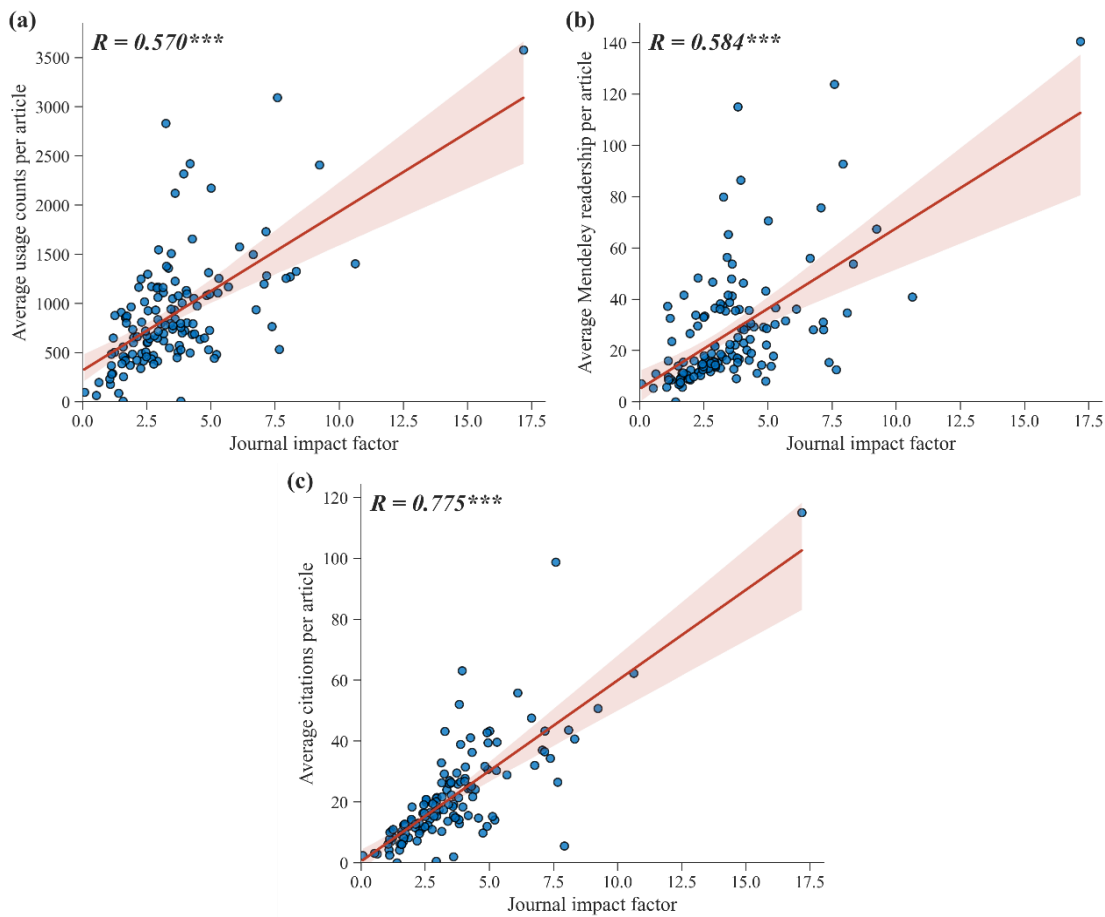

**Fig. 6** Spearman correlation between journal impact factor and (a) average usage counts, (b) Mendeley readership, and (c) citations at the journal-level. $N = 134$ unique journals

## 3.2 Regression analysis

To investigate the ability of early usage counts and early Mendeley readership to predict future citations of an article. Six linear regression models were constructed for both journal papers (see Table 2) and conference papers (see Table 3). Model 1 and Model 7 were developed using early usage data to predict citations, while Model 2 and Model 8 included additional control variables. Model 3 and Model 9 used early Mendeley data to forecast citations, with Model 4 and Model 10 incorporating control variables. Model 5 and Model 11 integrated early usage data and early Mendeley data to predict citations, while Model 6 and Model 12 added control variables.

The estimation results for journal papers indicate that after controlling for confounding variables, Model 6 has the best fit with an $R^2$ of 0.478. Thus, we focus on the estimation results of Model 6 in this study. In Model 6, early usage counts have a regression coefficient of 0.233, which is significant at the 0.001 level, indicating that an increase of one unit in early usage counts leads to a 0.233 unit increase in citations while holding all other variables constant. Similarly, early Mendeley readership has a regression coefficient of 0.529, which is significant at the 0.001 level, indicating that an increase of one unit in early Mendeley readership leads to a 0.529 unit increase in citations while all other variables remain constant.

**Table 2** Regression models for predicting citations by early usage counts or early Mendeley readership (journal papers)

| Variables | Predicting citations by early usage counts | | Predicting citations by early Mendeley readership | | Predicting citations by early usage counts and early Mendeley readership | |
|---|---|---|---|---|---|---|
| | Model 1 | Model 2 | Model 3 | Model 4 | Model 5 | Model 6 |
| Early usage counts | 0.499*** (0.007) | 0.461*** (0.007) | - | - | 0.237*** (0.007) | 0.233*** (0.007) |
| Early Mendeley readership | - | - | 0.656*** (0.006) | 0.635*** (0.006) | 0.541*** (0.007) | 0.529*** (0.007) |
| Funding | - | YES | - | YES | - | YES |
| International collaboration | - | YES | - | YES | - | YES |
| Open access | - | YES | - | YES | - | YES |
| JIF | - | YES | - | YES | - | YES |
| Adjusted $R^2$ | 0.249 | 0.277 | 0.430 | 0.437 | 0.473 | 0.478 |

*Standard errors are given in parentheses.

For conference papers, after controlling for confounding variables, Model 12 has the best fit with an $R^2$ of 0.427. Therefore, we analyze the estimation results of Model 12. In Model 12, the regression coefficient of early usage counts is 0.25, which is significant at the 0.001 level, indicating that an increase of one unit in early usage counts leads to a 0.25 unit increase in citations while all other variables remain constant. Similarly, the estimation result for early Mendeley readership is 0.540, which is significant at the 0.001 level, indicating that an increase of one unit in early Mendeley readership leads to a 0.540 unit increase in citations while all other variables remain constant.

Consistent estimation results for journal and conference papers provide empirical evidence for the predictive ability of early usage counts and early Mendeley readership for citations. Furthermore, the results suggest that differences in early Mendeley readership have a greater relationship with citations than early usage counts in the journal and conference papers. Additionally, further empirical research is necessary to investigate the mechanisms underlying the potential effects of early usage counts, early Mendeley readership, and citations. Thus, we continue to explore the mediating effects in the third part.

**Table 3** Regression models for predicting citations by early usage counts or early Mendeley readership (conference papers)

| Variables | Predicting citations by early usage counts | | Predicting citations by early Mendeley readership | | Predicting citations by early usage counts and early Mendeley readership | |
| --- | --- | --- | --- | --- | --- | --- |
| | Model 7 | Model 8 | Model 9 | Model 10 | Model 11 | Model 12 |
| Early usage counts | 0.380*** (0.003) | 0.379*** (0.003) | - | - | 0.249*** (0.003) | 0.250*** (0.003) |
| Early Mendeley readership | - | - | 0.606*** (0.003) | 0.601*** (0.003) | 0.547*** (0.003) | 0.540*** (0.003) |
| Funding | - | YES | - | YES | - | YES |
| International collaboration | - | YES | - | YES | - | YES |
| Open access | - | YES | - | YES | - | YES |
| Adjusted $R^2$ | 0.144 | 0.162 | 0.367 | 0.368 | 0.426 | 0.427 |

*Standard errors are given in parentheses.

### 3.3 Mediation Analysis

Assuming a rather stepwise literature retrieval process followed by scientists, which involves first browsing and accessing papers (e.g., by viewing the metadata or downloading the PDF version), then saving them to literature management software (e.g., like Mendeley), and then citing them, this study hypothesized that early Mendeley readership acts as a mediator between usage behavior and citation behavior. To test this assumption, we constructed a mediation model using SEM, the results of which are presented in Table 4. The findings demonstrate that early usage counts have a significant positive effect on citations for both journal and conference papers, as does early Mendeley readership. Moreover, the relationship between early usage counts and early Mendeley readership is also positive and significant. Although the first two relationships were reported in the regression analysis, they are included in Table 4 for the completeness of the SEM analysis. These results suggest that early Mendeley readership mediates the relationship between early usage counts and citations. Furthermore, we observed that the effect of early usage counts on early Mendeley readership is stronger for journal papers (coefficient of 0.484) than for conference papers (coefficient of 0.24).

**Table 4** Parameter estimation of SEM

| Variables | Journal papers | | Conference papers | |
| --- | --- | --- | --- | --- |
| | Early Mendeley | Citations | Early Mendeley | Citations |

|  | readership | | readership | |
|---|---|---|---|---|
| Early usage counts | 0.484*** | 0.233*** | 0.240*** | 0.250*** |
|  | (0.007) | (0.007) | (0.004) | (0.003) |
| Early Mendeley readership | - | 0.529*** | - | 0.540*** |
|  |  | (0.007) |  | (0.003) |
| Funding | YES | YES | YES | YES |
| International collaboration | YES | YES | YES | YES |
| Open access | YES | YES | YES | YES |
| JIF | YES | YES | - | - |

*Standard errors are given in parentheses.

In order to obtain a more accurate estimate of the mediating effect value of early Mendeley readership and to establish its confidence interval, a comprehensive approach was utilized in this study, which included the Delta, Sobel, and Monte Carlo simulation tests. The results, which are presented in Table 5 and Fig. 7, indicate that early Mendeley readership partially mediates the relationship between early usage counts and citations in both journal and conference papers. Specifically, for journal papers, the mediating effect value of early Mendeley readership on citations through early usage counts is 0.257, with a direct effect value of 0.234, and both are statistically significant at the 0.001 level. Similarly, for conference papers, the mediating effect value of early Mendeley readership on citations through early usage counts is 0.130, with a direct effect value of 0.251, and both are also statistically significant at the 0.001 level.

The finding that early Mendeley readership acts as a bridge between early usage counts and citations is in line with previous research. Some studies have shown that Mendeley readership is a useful indicator for predicting citation counts (Thelwall, 2018; Zahedi et al., 2017), which is also consistent with our regression results in Section 3.2. However, not all researchers use Mendeley for literature reading and management, and users typically only record the articles they plan to read or have already read and intend to cite (Mohammadi et al., 2016). There are still many researchers who save and cite articles through other literature management software, academic paper search platforms, or academic social networking sites. Consequently, early Mendeley readership only plays a partial mediating role.

**Table 5** Mediation effect test

|  | Journal papers | | | Conference papers | | |
|---|---|---|---|---|---|---|
|  | Delta | Sobel | Monte Carlo | Delta | Sobel | Monte Carlo |
| Indirect effect | 0.257 | 0.257 | 0.257 | 0.130 | 0.130 | 0.130 |
| SE | 0.004 | 0.004 | 0.004 | 0.002 | 0.002 | 0.002 |
| $z$-value | 60.205 | 60.950 | 60.864 | 66.314 | 65.548 | 65.808 |
| $p$-value | 0.000 | 0.000 | 0.000 | 0.000 | 0.000 | 0.000 |
| Conf. interval | [0.249, 0.266] | [0.249, 0.266] | [0.249, 0.266] | [0.126, 0.133] | [0.126, 0.133] | [0.126, 0.133] |

Furthermore, the mediating effect of early Mendeley readership on journal papers is more significant than that on conference papers, as conference papers rely more on the direct effect of early usage counts on citations. In journal papers, the value of indirect effects/direct effects is 1.1, while in conference papers, the value is only 0.516. This difference may be attributed to the fact that journal papers are typically longer and more complex in their research content, requiring scholars to spend more time reading and engaging with them, thus relying more on Mendeley for literature management and annotation recording. Conference papers are typically shorter in content; therefore, potential citers may not need to manage and engage with them through a reference manager but just view or download them directly from the publishers' website, thus explaining the lower mediating role of the Mendeley readership.

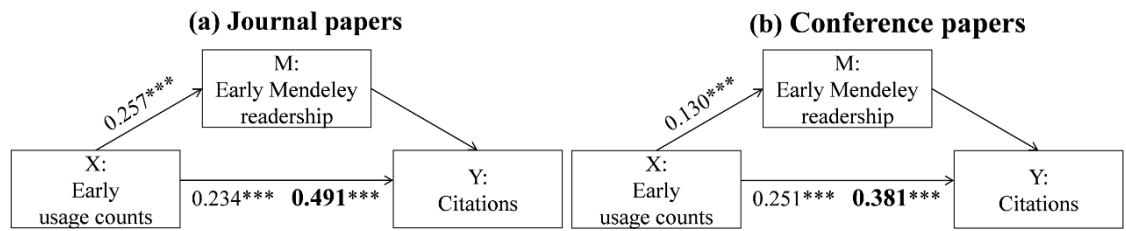

**Fig. 7** Path coefficients for the mediation effects models of (a) journal papers and (b) conference papers. The total effects are displayed in a larger font and in bold, and the direct and indirect effects are in smaller font sizes.

## 4 Conclusions

This study employs a multidimensional analysis, including descriptive, regression, and mediation analyses, of IEEE Xplore-published journal and conference papers from 2016 as data samples to compare the relationships among usage counts, Mendeley readership, and citations. This study represents a significant contribution to the field of scientometrics by extending the examination of the relationships among usage counts, Mendeley readership, and citation data beyond journal papers to conference papers. Furthermore, this study utilizes mediation techniques to offer novel insights into the relationships between early usage counts, early Mendeley readership, and citations. Additionally, we provide a valuable exploration of the IEEE Xplore database, which distinguishes this work from studies that rely on commonly used databases such as Web of Science and Scopus.

Results indicate no significant difference in the distribution patterns and accumulation rates between the two types of papers (journal and conference papers). Specifically, usage counts follow a pattern of first increasing and then decreasing, while Mendeley readership and citations follow a power-law distribution. Furthermore, the dissemination speed of the three indicators follows the order of "usage counts > Mendeley readership > citations." Regarding correlation, the correlation coefficient between these three indicators of conference papers (at around 0.56) is lower than that of journal papers (at around 0.70). In addition, funding, international collaboration, and open access have a positive relationship with all three metrics, with the only exception of the relationship between funding and the usage counts of conference papers. Notably, open access has the greatest association with the three metrics, suggesting the important role that open access may have in facilitating the process of accessing, reading, and eventually citing scientific papers. However, prior research suggests that only articles with immediate open access can benefit from these advantages, while delayed open access policies are ineffective in promoting knowledge dissemination in certain emerging and developing countries (X. Wang et al., 2015; Zhang et al., 2021).

The results of the regression analysis show that early Mendeley readership has better predictive power for citations compared to early usage counts, particularly in journal articles. Secondly, the combination of early usage counts and early Mendeley readership leads to the best predictive performance for citations. Furthermore, the mediation analysis demonstrates that early Mendeley readership partially mediates between early usage counts and citations in the journal and conference papers, with mediation values of 0.257 and 0.130, respectively. Additionally, compared to the direct effect of early usage counts on citations (0.234 and 0.251), conference papers rely more on this direct effect. This suggests that the form of engagement of users with the publications is similar for both journal and conference papers, although with some differences, with the usage counts of conference papers being more strongly associated with citations without the mediation of readership.

This study presents several avenues for further research. First, additional altmetric indicators, like tweets, news media mentions, or Wikipedia citations, could also be included in a path analysis to better understand the mechanisms underlying their relationship with eventual citation counts and among themselves. Secondly, causality could be further explored to move beyond correlational analyses, which would enable a more accurate interpretation of the actual effects of different usage and readership events on citation counts. Finally, finer-grained time-series data, such as monthly data, could be constructed for early usage counts and early Mendeley readership, and time-series forecasting methods combined with machine learning methods could be used to predict future citations with greater precision.

## Acknowledgements

This study is partially supported by the National Natural Science Foundation of China (71974029). Rodrigo Costas is partially funded by the South African DSI-NRF Centre of Excellence in Scientometrics and Science, Technology and Innovation Policy (SciSTIP). Wencan Tian is financially supported by the China Scholarship Council (202106060134). Zhichao Fang is funded by the Scientific Research Funding of Renmin University of China (No. 23XNF037).

## Conflict of interest

One of the authors (Rodrigo Costas) is a member of the Distinguished Reviewers Board of the journal *Scientometrics*.

## Data and code availability

Data and codes can be accessed from this URL:

https://github.com/Tianwencan/IEEE_usage

# Appendix

**Table 6** Descriptive statistics for usage counts, Mendeley readership and citations

| Publication Type | Metrics | N | Min | Max | Mean | Std |
|---|---|---|---|---|---|---|
| Journal | Usage counts | 16,799 | 6 | 140,544 | 851.835 | 2163.630 |
| | Mendeley readership | 16,799 | 0 | 5,256 | 20.383 | 62.112 |
| | Citations | 16,799 | 0 | 2,103 | 19.646 | 42.312 |
| Conference | Usage counts | 98,773 | 7 | 18,103 | 248.755 | 331.802 |
| Mendeley readership | 98,773 | 0 | 3,072 | 10.756 | 29.124 |
| Citations | 98,773 | 0 | 922 | 4.412 | 14.540 |

**Table 7** The specific information statistics for Fig. 5

| Indicators | Metrics | N (Yes) | N (No) | Value (Yes) | Value (No) | F value | P value |
|---|---|---|---|---|---|---|---|
| Funding (journal) | Usage counts | 10,222 | 6,577 | 898.99 | 778.55 | 1,241 | p<0.001 |
| | Mendeley readership | 10,222 | 6,577 | 21.62 | 18.46 | 1,041 | p<0.01 |
| | Citations | 10,222 | 6,577 | 22.04 | 15.92 | 8,424 | p<0.001 |
| International collaboration (journal) | Usage counts | 4,899 | 11,900 | 1040.06 | 774.35 | 5,250 | p<0.001 |
| | Mendeley readership | 4,899 | 11,900 | 24.81 | 18.56 | 3,522 | p<0.001 |
| | Citations | 4,899 | 11,900 | 25.67 | 17.17 | 14,118 | p<0.001 |
| Open access (journal) | Usage counts | 4,535 | 12,264 | 1218.90 | 716.10 | 18,071 | p<0.001 |
| | Mendeley readership | 45,35 | 12,264 | 29.82 | 16.89 | 14,453 | p<0.001 |
| | Citations | 45,35 | 12,264 | 25.59 | 17.45 | 12,355 | p<0.001 |
| Funding (conference) | Usage counts | 37,695 | 61,078 | 246.41 | 250.20 | 304 | *No.sig* |
| | Mendeley readership | 37,695 | 61,078 | 12.40 | 9.74 | 19,416 | p<0.001 |
| | Citations | 37,695 | 61,078 | 5.27 | 3.88 | 21,286 | p<0.001 |
| International collaboration (conference) | Usage counts | 15,450 | 83,323 | 264.38 | 245.86 | 4,063 | p<0.001 |
| | Mendeley readership | 15,450 | 83,323 | 14.29 | 10.10 | 27,095 | p<0.001 |
| | Citations | 15,450 | 83,323 | 6.15 | 4.09 | 26,216 | p<0.001 |
| Open access (conference) | Usage counts | 8,340 | 90,433 | 295.53 | 244.44 | 18,137 | p<0.001 |
| | Mendeley readership | 8,340 | 90,433 | 26.31 | 9.32 | 266,682 | p<0.001 |
| | Citations | 8,340 | 90,433 | 10.89 | 3.82 | 184,156 | p<0.001 |

* "Yes" means that articles in this group were funded by grants or developed international collaborations or were open access. "No" means the opposite.